\documentclass[aip,preprint]{revtex4-1}
\pdfoutput=1
\usepackage{graphicx}
\usepackage{amsmath}
\usepackage{amsfonts}
\usepackage{amssymb}
\usepackage{bm}
\usepackage{color}
\usepackage{multirow}

\usepackage[framemethod=TikZ]{mdframed}
\mdfdefinestyle{CommentFrame}{%
    linecolor=black,
    outerlinewidth=1pt,
    roundcorner=5pt,
    innertopmargin=0.5\baselineskip,
    innerbottommargin=\baselineskip,
    innerrightmargin=0pt,
    innerleftmargin=0pt,
    backgroundcolor=gray!20!white,
    nobreak=true}

\begin{document}
\preprint{APS/123-QED}

\title{Mapping continuous potentials to discrete forms}

\author{Chris Thomson}
\affiliation{
  School of Engineering, University of Aberdeen,
  Fraser Noble Building, Kings College, Aberdeen AB24 3UE, UK}

\author{Leo Lue}
\affiliation{
  Department of Chemical and Process Engineering, University of Strathclyde,
  James Weir Building, 75 Montrose Street, Glasgow G1 IXJ, UK}

\author{Marcus N. Bannerman}
\affiliation{
  School of Engineering, University of Aberdeen,
  Fraser Noble Building, Kings College, Aberdeen AB24 3UE, UK}
\email{m.campbellbannerman@abdn.ac.uk}

\date{\today}

\begin{abstract}
  The optimal conversion of a continuous inter-particle potential to a
  discrete equivalent is considered here. Existing and novel
  algorithms are evaluated to determine the best technique for
  creating accurate discrete forms using the minimum number of
  discontinuities. This allows the event-driven molecular dynamics
  technique to be efficiently applied to the wide range of continuous
  force models available in the literature, and facilitates a direct
  comparison of event-driven and time-driven molecular dynamics. The
  performance of the proposed conversion techniques are evaluated
  through application to the Lennard-Jones model. A surprising linear
  dependence of the computational cost on the number of
  discontinuities is found, allowing accuracy to be traded for speed
  in a controlled manner. Excellent agreement is found for static and
  dynamic properties using a relatively low number of
  discontinuities. For the Lennard-Jones potential, the optimized
  discrete form outperforms the original continuous form at gas
  densities but is significantly slower at higher densities.
\end{abstract}

\pacs{Valid PACS appear here}


\maketitle
%%%%%%%%%%%%%%%%%%%%%%%%%%%%%%%%%%%%%%%%%%%%%%%%%%%%%%%%%%%%%%%%%%%%%%%%%%%%%%%%
%%%%%%%%%%%%%%%%%%%%%%%%%%%%%%%%%%%%%%%%%%%%%%%%%%%%%%%%%%%%%%%%%%%%%%%%%%%%%%%%
%%%%%%%%%%%%%%%%%%%%%%%%%%%%%%%%%%%%%%%%%%%%%%%%%%%%%%%%%%%%%%%%%%%%%%%%%%%%%%%%
\section{Introduction}

Particle simulation techniques are now over 50 years
old~\cite{Alder_Wainwright_1957} and have become a vital tool in
exploring natural processes at all scales.  Molecular dynamics,
granular dynamics~\cite{Poschel_Schwager_2005}, dissipative particle
dynamics, and even smooth particle
hydrodynamics~\cite{Gingold_Monaghan_1977} algorithms are all
fundamentally identical.  They each attempt to solve classical
equations of motion for a large number of particles. In such models,
conservative interactions between particles are typically defined
through a pairwise additive inter-particle potential $\Phi(r)$, where
$r$ is the distance between the particles.  The force ${\bf F}_{ij}$
acting on particle $i$ due to particle $j$ is given by
\begin{align*}
  {\bf F}_{ij} 
  = -\frac{\partial}{\partial{\bf r}_i} \Phi(|{\bf r}_i-{\bf r}_j|)
\end{align*}
where ${\bf r}_i$ is the position of particle $i$, and ${\bf r}_j$ is
the position of particle $j$.

\begin{figure}
  \centering
  \includegraphics[clip,width=\columnwidth]{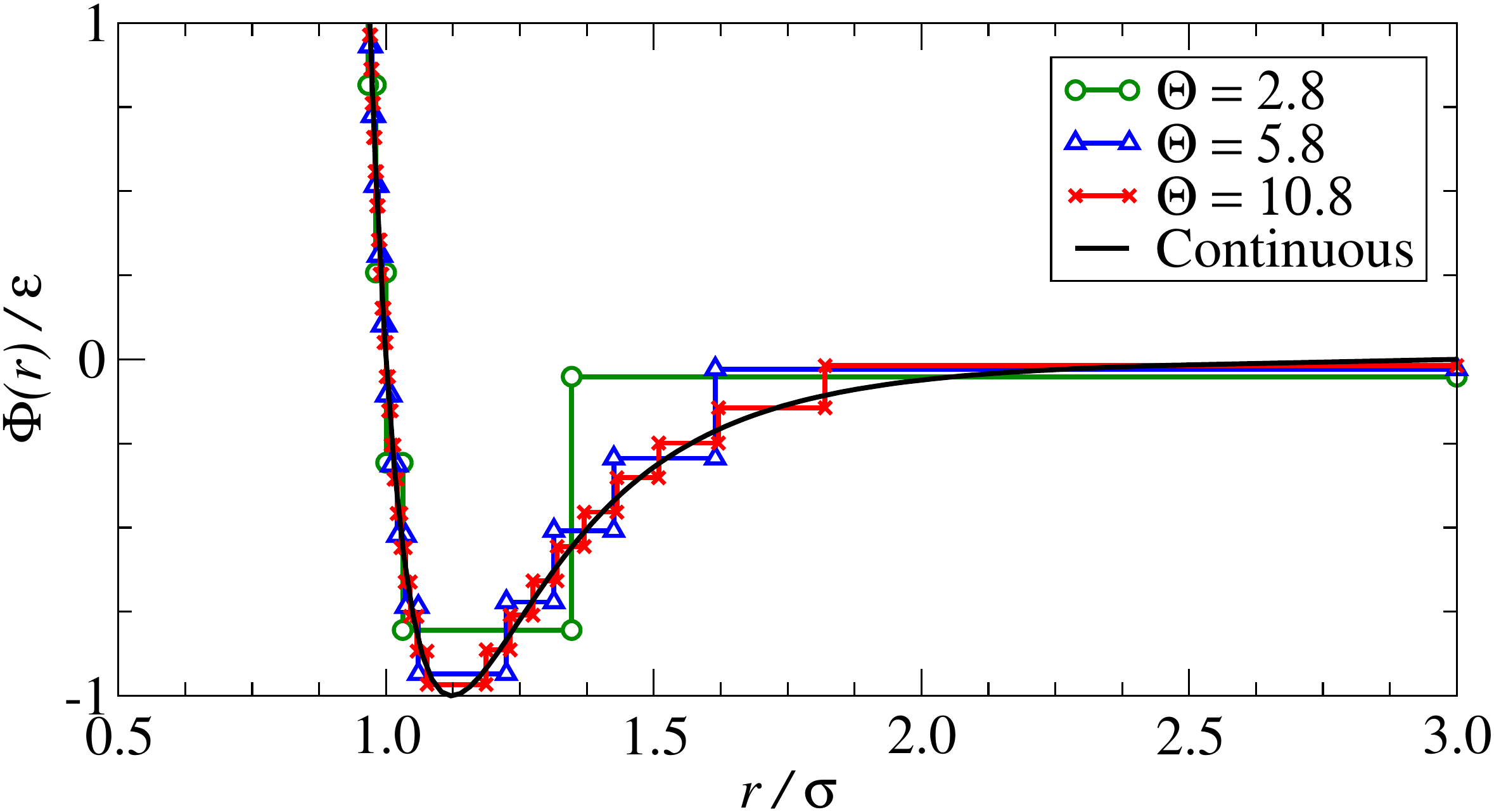}
  \caption{\label{fig:LJstepped}%
    A comparison of the continuous Lennard-Jones potential (solid) and
    three stepped approximations created using
    Eq.~\eqref{eq:deltaustepping} for step placement and
    Eq.~\eqref{eq:VolumeEnergyStepping} for the step energies.}
\end{figure}

There are two broad categories of inter-particle potentials: continuous
and discrete.  For continuous potentials, the interaction energy is a
continuous function of the particle positions. The Lennard-Jones
potential is a classic example of a continuous potential:
\begin{align*}
  \Phi^{\textrm{LJ}}(r) =
  4\,\varepsilon\left[
    \left(\frac{\sigma}{r}\right)^{12}
      - \left(\frac{\sigma}{r}\right)^{6}
  \right]
\end{align*}
where $r$ is the distance between the two particles, $\varepsilon$ is
the minimum interaction energy, and $\sigma$ is the separation
distance corresponding to zero interaction energy. In discontinuous
(also known as ``stepped'' or
``terraced''~\cite{vanZon_Schofield_2008}) potentials, the interaction
potential changes only at discrete locations and a functional
definition is difficult. An illustration of the two forms of the
Lennard-Jones potential is given in Fig.~\ref{fig:LJstepped}.

Continuous potentials are popular as the finite-difference algorithms
used to simulate them are well-known~\cite{Haile_1997} and it is
straightforward to implement physical scaling laws into the model
potential. For example, the $r^{-6}$ term in the Lennard-Jones
potential was selected to match the known scaling of molecular
dispersion forces. Discontinuous potentials on the other hand are
typically reported as a table of discontinuity locations and
energies~\cite{Chapela_etal_1989}.  Although these two classes of
potentials are distinct, it is clear that they may be made equivalent,
provided a sufficient number of discontinuities or steps are used, as
illustrated in Fig.~\ref{fig:LJstepped}.  The optimal number,
location, and energetic change of the discontinuities for an accurate
representation of a continuous potential is not well understood and is
the subject of this paper.

%%%
%%%
%%%
%%%

The motivation for this study is to understand the equivalence between
the two approaches and to allow conversion between them. Continuous
potentials are prevalent in the simulation literature, beginning with
the first simulations of Lennard-Jones systems by Verlet in
1967~\cite{Verlet_1967} to the complex many-body potentials used for
biological systems today~\cite{Ponder_Case_2003,CHARMMFF}. Discrete
potentials are equally as popular due to their amenability to
theoretical analysis, and are at the heart of thermodynamic
perturbation theory (TPT)~\cite{Barker_Henderson_1967} and kinetic
theory~\cite{Chapman_Cowling_1991}.
However, there has not been the same explosion of molecular force
fields and software tools as for continuous potentials (e.g.,
GROMACS~\cite{VanderSpoel_etal_2005} and
ESPResSo~\cite{Limbac_etal_2006}).  This is even more surprising given
that the very first particle simulations were carried out using a
discrete potential~\cite{Alder_Wainwright_1957}, almost ten years
before Verlet's simulations.

It is only relatively recently that fine-tuned discrete potentials for
detailed, atomistic simulations have started to appear; these include
force fields for a broad range of compounds, including
hydrocarbons~\cite{Cui_Elliott_2002} and
fluorocarbons~\cite{Sans_Elliott_2008}, organic
acids~\cite{Vahid_etal_2008}, esters, ketones and other organic
compounds~\cite{Baskaya_etal_2005,Hassan_etal_2012},
phospholipids~\cite{Curtis_Hall_2013}, and peptides and
proteins~\cite{Nguyen_Hall_2004}.  The use of TPT has even allowed
rapid and direct fitting of discrete potentials to experimental
data~\cite{Cui_Elliott_2002,Unlu_etal_2004,dosRamos_etal_2009}.  In
addition, standard simulation packages for event-driven molecular
dynamics have also begun to appear~\cite{Bannerman_etal_2011}.
The strong theoretical frameworks and stable simulation algorithms
make discrete potentials an attractive alternative to continuous
potentials; therefore, it is desirable to have a mechanism to map
existing continuous potentials into discrete forms.

This mapping must be optimized in the sense that it must use the
smallest number of discontinuities to reduce the complexity of the
converted potential and to minimize the computational cost of
simulation.  Chapela et al.~\cite{Chapela_etal_1989} was the first to
attempt to represent the continuous Lennard-Jones potential by an
equivalent discrete form.  This mapping was optimized ``by hand'' to
reproduce the thermodynamic properties at one state point, but more
recent work has focused on using regular step
placement~\cite{Chapela_etal_2010,Chapela_etal_2013} and algorithms to
determine the step
energies~\cite{Chapela_etal_2010,Ucyigitler_etal_2012,Chapela_etal_2013,Torres-Arenas_etal_2010}
to partially automate the process.  Algorithms for directly specifying
both the location and energy changes of discontinuities from
underlying continuous potentials have also been
presented~\cite{vanZon_Schofield_2008,Torres-Arenas_etal_2010}
allowing a convenient implementation of arbitrary potentials in
event-driven dynamics; however, the optimization of direct conversion
is yet to be explored.  Recently, there has been an attempt to replace
the soft interactions of continuous potentials entirely with collision
dynamics at low densities~\cite{Muller_Poschel_2013} but this approach
is restricted to low density systems.

In this work, the mapping of a continuous potential to a discrete form
is investigated using the Lennard-Jones potential. In the following
section, the placement of discontinuities and allocation of step
energies is discussed before the methods are evaluated in
Sec.~\ref{sec:evalulation}. The most efficient mapping scheme is then
evaluated for a range of thermodynamic and transport properties in
Sec.~\ref{sec:performance}. A comparison between time-stepping and
event-driven simulation is performed in Sec.~\ref{sec:compcost}.
Finally, the conclusions of the paper are presented in
Sec.~\ref{sec:conclusions}.

%%%%%%%%%%%%%%%%%%%%%%%%%%%%%%%%%%%%%%%%%%%%%%%%%%%%%%%%%%%%%%%%%%%%%%%%%%%%%%%%
%%%%%%%%%%%%%%%%%%%%%%%%%%%%%%%%%%%%%%%%%%%%%%%%%%%%%%%%%%%%%%%%%%%%%%%%%%%%%%%%
\section{Discretization of the potential}

The primary aim of this work is to develop an algorithm to convert a
continuous potential to an optimal discrete form: one that provides an
accurate approximation of the original continuous potential and can be
simulated at a minimal computational cost.  As the computational cost
of an event-driven simulation is roughly proportional to the number of
discontinuities encountered by the particles, it is vital that the
number of discontinuities or steps used to achieve a set level of
accuracy is minimized.

The location of a single step $i$ in a spherically-symmetric discrete
potential is specified by the segment $[r_{i+1},r_{i}]$ between the
$i$th and $i+1$th discontinuities.  The discontinuities, located at
$r_i$ and $r_{i+1}$, bounding each step are also the limits of the
neighboring steps.  A simplification made in this work is to require
that each step is directly representative of the segment of the
continuous potential lying within the same limits $[r_{i+1},r_{i}]$.
This allows the task of discretizing the potential to be split into
two smaller tasks: the optimal placement of discontinuities and the
determination of an effective step energy for a segment of the
continuous potential.

It is common to accelerate molecular dynamics calculations by
truncating the interaction potential at a cut-off radius
$r_{\text{cutoff}}$, thus requiring only local particle pairings to be
considered in force calculations.  Typically in simulations of
continuous potentials, the potential is also shifted to eliminate the
discontinuity at the cutoff in order to avoid the presence of
impulsive forces.  For example, the truncated, shifted Lennard-Jones
potential is given by
\begin{align}
\Phi(r) = 
  \begin{cases}
    \Phi^{LJ}(r) -\Phi^{LJ}(r_{\text{cutoff}}) 
      & \text{if $r\le r_{\text{cutoff}}$}\\
    0 & \text{if $r>r_{\text{cutoff}}$}
  \end{cases}
.
\end{align}

As each step of the discontinuous potential represents a segment of
the original continuous potential, the first discontinuity is defined
to lie at the cutoff radius (i.e.\ $r_1=r_{\text{cutoff}}$), while all
other discontinuities lie within in the region
$r\in(0,r_{\text{cutoff}})$.  It is tempting to also define an inner
hard-core radius of the stepped potential using one of the available
methods (e.g., see Ref.~\citenum{Barker_Henderson_1967b}); however,
this would require each step energy to somehow compensate for the
overly repulsive core, inextricably linking step placement and energy
once again.  The available methods for placing discontinuities are
reviewed in the next section before the algorithms used to generate
representative step energies are discussed.

%%%%%%%%%%%%%%%%%%%%%%%%%%%%%%%%%%%%%%%%%%%%%%%%%%%%%%%%%%%%%%%%%%%%%%%%%%%%%%%%
\subsection{Location of Discontinuities}

The simplest approach to place the discontinuities of a discrete
potential is to divide the region $r\in(0,r_{\text{cutoff}})$ into a
number of steps of equal width $\Delta r$~\cite{Chapela_etal_2010}.
\begin{align}\label{eq:deltarstepping}
  r_{i,\Delta r} = r_{\text{cutoff}} - (i-1) \Delta r
\end{align}
The total number of discontinuities/steps in the potential (including
the cutoff) is given by $\left\lfloor r_{\text{cutoff}}/\Delta
  r\right\rfloor+1$.  

It is not immediately clear that a uniform radial placement of the
steps is the natural choice for a spherical potential.  An alternative
choice is to fix the volume $\Delta v$ bounded by each step of the
potential.  In this case, each step location is determined using the
following recursive expression
\begin{align}\label{eq:deltavstepping}
  r_{i,\Delta v} &= \left(r^3_{i-1,\Delta v}-\frac{3\,\Delta
      v}{4\,\pi}\right)^{1/3}
\end{align}
The total number of discontinuities in the potential is then
$\left\lfloor 4\,\pi\,r_{\text{cutoff}}^3/3\,\Delta v\right\rfloor+1$.

The primary disadvantage of the approaches outlined above is that they
do not attempt to adapt the step locations according to the behavior
of the potential.  It is likely that the performance of both
algorithms is particularly sensitive to the configuration of the steps
near the minimum of the potential where the interaction energy
changes rapidly.

It has also been
proposed~\cite{vanZon_Schofield_2008,Chapela_etal_2013} to discretize
continuous potentials by placing discontinuities at fixed intervals of
interaction energy $\Delta \Phi$. This approach allows a controlled
resolution of the potential, while balancing the contribution of each
step and allows a straightforward extension to asymmetric
potentials. The locations of the discontinuities are the ordered
solutions to the following set of equations
\begin{align}\label{eq:deltaustepping}
  \Phi(r) &= j\,\Delta \Phi & j\in\mathbb{Z}
\end{align}
The application of Eq.~\eqref{eq:deltaustepping} to the shifted,
truncated Lennard-Jones potential results in an infinite number of
steps due to the singularity at $r=0$. In practice, the high-energy
steps are inaccessible and only a small number need to be computed
during the simulation.

Before these approaches can be evaluated, a technique for determining
the step energies must be selected.  This is discussed in the
following section.

%%%%%%%%%%%%%%%%%%%%%%%%%%%%%%%%%%%%%%%%%%%%%%%%%%%%%%%%%%%%%%%%%%%%%%%%%%%%%%%%
\subsection{Step Energy}

With the location of each step defined through one of the above
algorithms, an algorithm for determining the effective energy of a
segment of the potential is required. In the limit of a large number
of discontinuities/small segments, the original continuous potential
must be recovered.  Chapela et al.~\cite{Chapela_etal_2010} have
evaluated three approaches based on point sampling of the continuous
potential.
\begin{align}\label{eq:LeftEnergyStepping}
  \Phi_i^{Left} &= \Phi\left(r_{i+1}\right) \\
  \label{eq:MidEnergyStepping}
  \Phi_i^{Mid} &= \Phi\left(\frac{r_{i}+r_{i+1}}{2}\right)\\
  \label{eq:RightEnergyStepping}
  \Phi_i^{Right} &= \Phi\left(r_{i}\right)
,
\end{align}
where $\Phi_i$ is the energy of step $i$ over the region
$[r_{i+1},r_{i}]$ of the discontinuous potential. Chapela et
al.~\cite{Chapela_etal_2010} report that mid-point sampling
($\Phi_i^{Mid}$) of the underlying continuous potential is the most
effective at reproducing the original behavior of the Lennard-Jones
potential, whereas left sampling is more appropriate for the Yukawa
potential. It is easy to define other methods of point sampling, such
as the minimum edge energy used by van Zon and
Schofield~\cite{vanZon_Schofield_2008}.  A logical choice is to sample
the potential at the distance which divides the step into two equal
volumes.
\begin{align}\label{eq:MidVolEnergyStepping}
\Phi_i^{Mid~Vol} &= \Phi\left(\left[\frac{r_{i}^3+r_{i+1}^3}{2}\right]^{1/3}\right)
.
\end{align}

It is also possible to define alternative approaches which do not rely
on point-sampling, such as an equal area
approach~\cite{Chapela_etal_2013}; however, it is more desirable to
directly match the thermodynamic properties of the converted
potentials.  Unfortunately, matching properties, such as the pressure,
would require the use of an accurate free energy, which is typically
unavailable.  
Successful attempts have been made to adjust stepped potentials to
directly match the predictions of the TPT to experimental data for a
range of thermodynamic properties~\cite{Ucyigitler_etal_2012};
however, the resulting equation of state is still approximate and the
expressions are rather complicated.  In this work, the focus is on
directly reproducing the properties of the continuous potential
system. One simple approach is to use the lowest order density
correction to both the pressure and free-energy, given by the second
virial coefficient, which is directly calculated from the interaction
potential.  The contribution of a segment of the potential to the
second virial may be calculated as follows
\begin{align}
  B_2(r_i,\ r_{i+1},\ T) &= -2\,\pi\int_{r_{i+1}}^{r_{i}}
    \left(e^{-\beta\Phi(r)}-1\right)r^2\,{\rm d}r
\end{align}
where $\beta=1/(k_BT)$, $T$ is the absolute temperature of the system,
and $k_B$ is the Boltzmann constant.  The energy of the step can then
be set to match the contribution to the second virial coefficient for
the corresponding segment of the continuous potential, using the
following expression
\begin{align}\label{eq:VirialEnergyStepping}
  \Phi_i^{virial}(T) &=
  -k_B\,T\ln\left(\frac{3}{r_{i}^3-r_{i+1}^3}\int_{r_{i+1}}^{r_{i}}r^2
  \,e^{-\beta\Phi(r)}\,{\rm d}r\right)
.
\end{align}
Application of this algorithm leads to excellent agreement at low
densities for the pressure; however, the algorithm has a cumbersome
dependence on the temperature, which may not be available {\em a
  priori} (e.g., in the microcanonical ensemble). In the
high-temperature limit, equating the virial contribution reduces to
taking a volume average of the energy within a step:
\begin{align}\label{eq:VolumeEnergyStepping}
  \Phi_i^{Volume} &= \frac{3}{r_{i}^3-r_{i+1}^3}\int_{r_{i+1}}^{r_{i}}
  \Phi \left(r\right) r^2\, {\rm d}r
.
\end{align}
This indicates that a volume-averaged approach will also yield a good
reproduction of the pressure near the ideal gas limit of
high-temperature and low-density.  It should be noted that both virial
and volume-averaging approaches will set an infinite energy for the
innermost step of the Lennard-Jones potential due to the singularity
at $r=0$. This can have a dramatic effect on the potential as the step
placement algorithms in Eqs.~\eqref{eq:deltavstepping} and
\eqref{eq:deltaustepping} use a finite number of steps to represent
the repulsive core.

%%%%%%%%%%%%%%%%%%%%%%%%%%%%%%%%%%%%%%%%%%%%%%%%%%%%%%%%%%%%%%%%%%%%%%%%%%%%%%%%
%%%%%%%%%%%%%%%%%%%%%%%%%%%%%%%%%%%%%%%%%%%%%%%%%%%%%%%%%%%%%%%%%%%%%%%%%%%%%%%%
\section{Comparison of mapping procedures
\label{sec:evalulation} }

To compare the various methods for mapping potentials, molecular
dynamics simulations of $N=1372$ discontinuous and continuous
Lennard-Jones particles with $r_{\text{cutoff}}=3\sigma$ were
performed over a range of densities, $\rho=N/V$, and temperatures,
$k_B\,T$.  To collect thermodynamic properties, each simulation was
run for $20\,(m\sigma^2/\varepsilon)^{1/2}$ for equilibration before
five production runs of $30\,(m\sigma^2/\varepsilon)^{1/2}$ were used
to collect averages and obtain estimates of the uncertainty.
Dynamical properties were collected using three runs, each
$10^3\,(m\sigma^2/\varepsilon)^{1/2}$ in duration.  Averages are
reported here with error bars corresponding to the standard deviation
between runs. Simulations for the continuous truncated, shifted
Lennard-Jones potential were performed using the
ESPResSo~\cite{Limbac_etal_2006} package with a time step of
$0.002\,(m\sigma^2/\varepsilon)^{1/2}$ and a Langevin thermostat with
a friction parameter of $1\,(\varepsilon/m\sigma^2)^{1/2}$.  Discrete
potential simulations were performed using the
DynamO~\cite{Bannerman_etal_2011} package with an Andersen thermostat
controlled to $5$\% of the overall events.  During the collection of
dynamical properties, the thermostat is disabled after the
equilibration period and the temperature is monitored to ensure it
remains within $2$\% of the set value.

The mapping procedures must be evaluated on a basis of accuracy as a
function of computational cost.  As event-driven simulators process
events at a constant rate, the computational cost is proportional to
the number of events that must be processed per unit of simulation
time.  Each additional discontinuity within the potential will
generate additional events provided particle pairs can access it;
therefore, the computational cost is dominated by the number of
discontinuities in the well of the potential.  A straightforward
parameter for the order of approximation of the potential, $\Theta$,
can then be defined for each step placement algorithm as follows:
\begin{align*}
\Theta &= 1+\left\{
  \begin{array}{ll}
    \displaystyle
    \left(r_{\text{cutoff}}-r_{\text{min}}\right)/\Delta r 
      & \text{\ for Eq.~\eqref{eq:deltarstepping}}\\
    \displaystyle
    4\,\pi\left(r_{\text{cutoff}}^3-r_{\text{min}}^{3}\right)/3\,\Delta v
      & \text{\ for Eq.~\eqref{eq:deltavstepping}} \\
    \displaystyle
    -\Phi(r_{\text{min}})/\Delta\Phi 
      & \text{\ for Eq.~\eqref{eq:deltaustepping}}\\
  \end{array}
\right.
\end{align*}
where $r_{\text{min}}=2^{1/6}\,\sigma$ is the location of the minimum
of the Lennard-Jones potential.  The parameter $\Theta$ is continuous,
and as $\Theta\to\infty$, the continuous Lennard-Jones potential is
recovered. The integer part of $\Theta$ corresponds to the number of
discontinuities in the attractive section of the potential and at whole
integer values of $\Theta$, a discontinuity is placed at the minimum
of the potential.

%%%%%%%%%%%%%%%%%%%%%%%%%%%%%%%%%%%%%%%%%%%%%%%%%%%%%%%%%%%%%%%%%%%%%%%%%%%%%%%%
\subsection{Placement of the discontinuities}

\begin{figure}
  \centering
  \includegraphics[width=\columnwidth,clip]{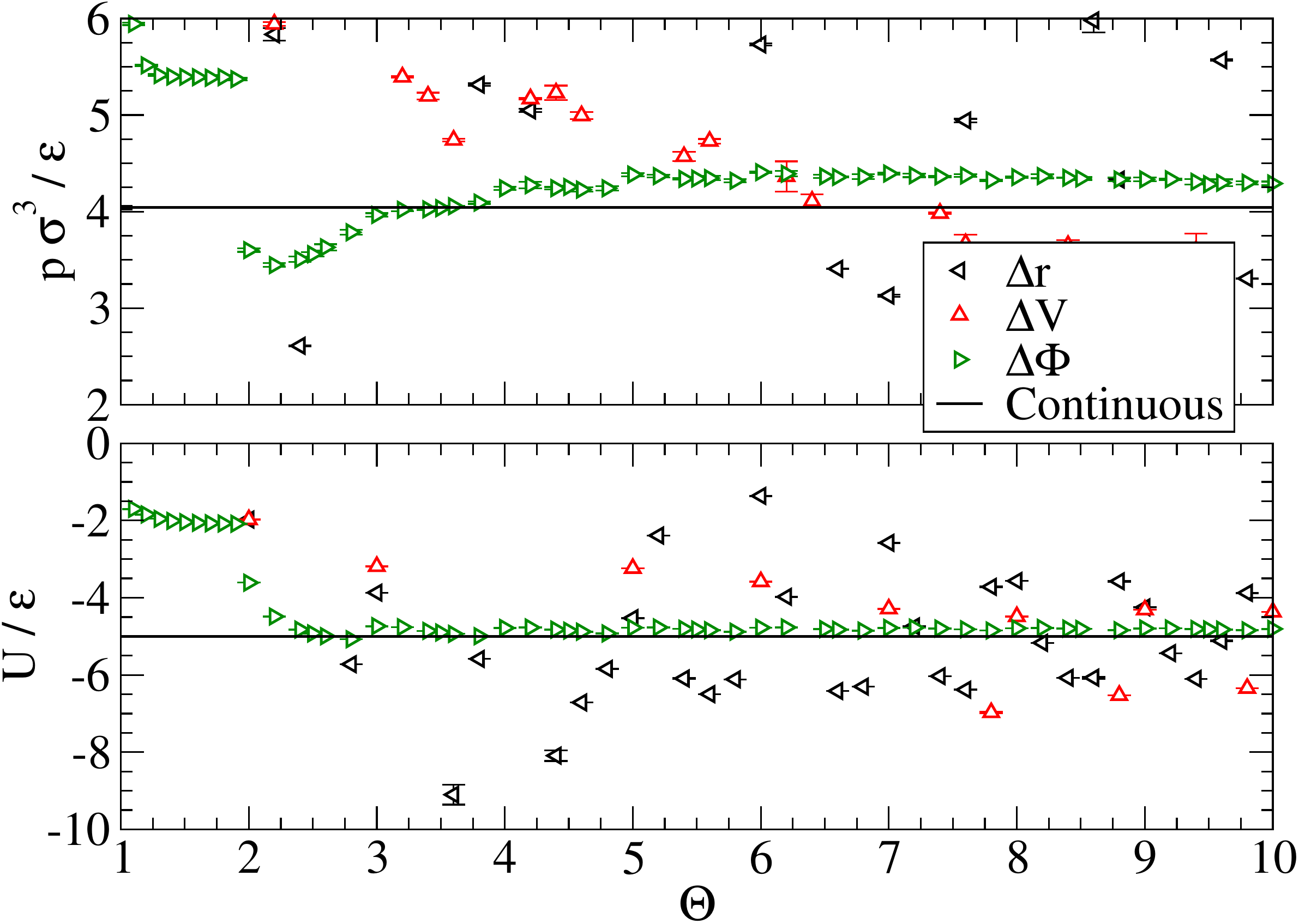}
  \caption{\label{fig:placementcomparision}Pressure, $p$, and internal
    energy, $U$, for a super-critical LJ liquid
    ($k_B\,T/\varepsilon=1.3$, $\rho\sigma^3=0.85$) as a function of the
    number of attractive steps, $\Theta$, in the
    potential. The solid line indicates the continuous potential
    result. Each set of data points correspond to a different
    algorithm for placing the steps whereas the $\Phi^{mid}_i$
    algorithm is used to set the step energy in all cases.}
\end{figure}

The methods for placing the discontinuities of a discrete potential
(Eq.~\eqref{eq:deltarstepping}--\eqref{eq:deltaustepping}) are
evaluated first. An example comparison of the calculated pressure and
internal energy for a high-density super-critical state point using
the mid-point sampling algorithm (Eq.~\eqref{eq:MidEnergyStepping}) to
set the step energies is given in Fig.~\ref{fig:placementcomparision}.
A temperature of $k_B\,T/\varepsilon=1.3$ is used in these simulations
as it is well-above the critical temperature
$k_BT_c/\varepsilon\approx1.15$ of the $r_{\text{cutoff}}=3\,\sigma$
system~\cite{Smit_1992} to avoid entering the two-phase region.  It is
clear from this comparison alone that the algorithm using fixed energy
intervals ($\Delta\Phi$, Eq.~\eqref{eq:deltaustepping}) is superior to
the other approaches. The algorithm demonstrates rapid convergence to
the result obtained from the continuous potential simulations once
$\Theta>2$, and allows a controlled approximation of the continuous
potential. The slight offset in the pressure at higher values of
$\Theta$ is later shown to be an artifact of using the mid-point
sampling method for the step energy.  The radial placement algorithm
(Eq.~\eqref{eq:deltarstepping}, $\Delta r$) appears to converge
towards the continuous potential result for the internal energy but
requires a large number of steps and its performance is erratic. The
radial and volumetric placement algorithms are particularly sensitive
to the order of approximation.  For example, the volumetric algorithm
is only accurate for the internal energy at integer values of $\Theta$
which correspond to a discontinuity placed at the minimum of the
potential. This highlights the importance of a good approximation of
the potential in this region at high densities.

Further simulations have been carried out using all step-energy
algorithms over a range of densities and temperatures and are in
qualitative agreement with the trends outlined in
Fig.~\ref{fig:placementcomparision}; therefore, it is clear that the
fixed energy interval algorithm given in Eq.~\eqref{eq:deltaustepping}
is the most appropriate as it is the only approach which allows a
controlled approximation of the stepped potential.  A full review of
the available step energy algorithms using the fixed energy interval
algorithm is performed in the following subsection.

%%%
%%%
%%%
%%%

%%%%%%%%%%%%%%%%%%%%%%%%%%%%%%%%%%%%%%%%%%%%%%%%%%%%%%%%%%%%%%%%%%%%%%%%%%%%%%%%
\subsection{Step energy}

\begin{figure}
  \centering
  \includegraphics[width=\columnwidth,clip]{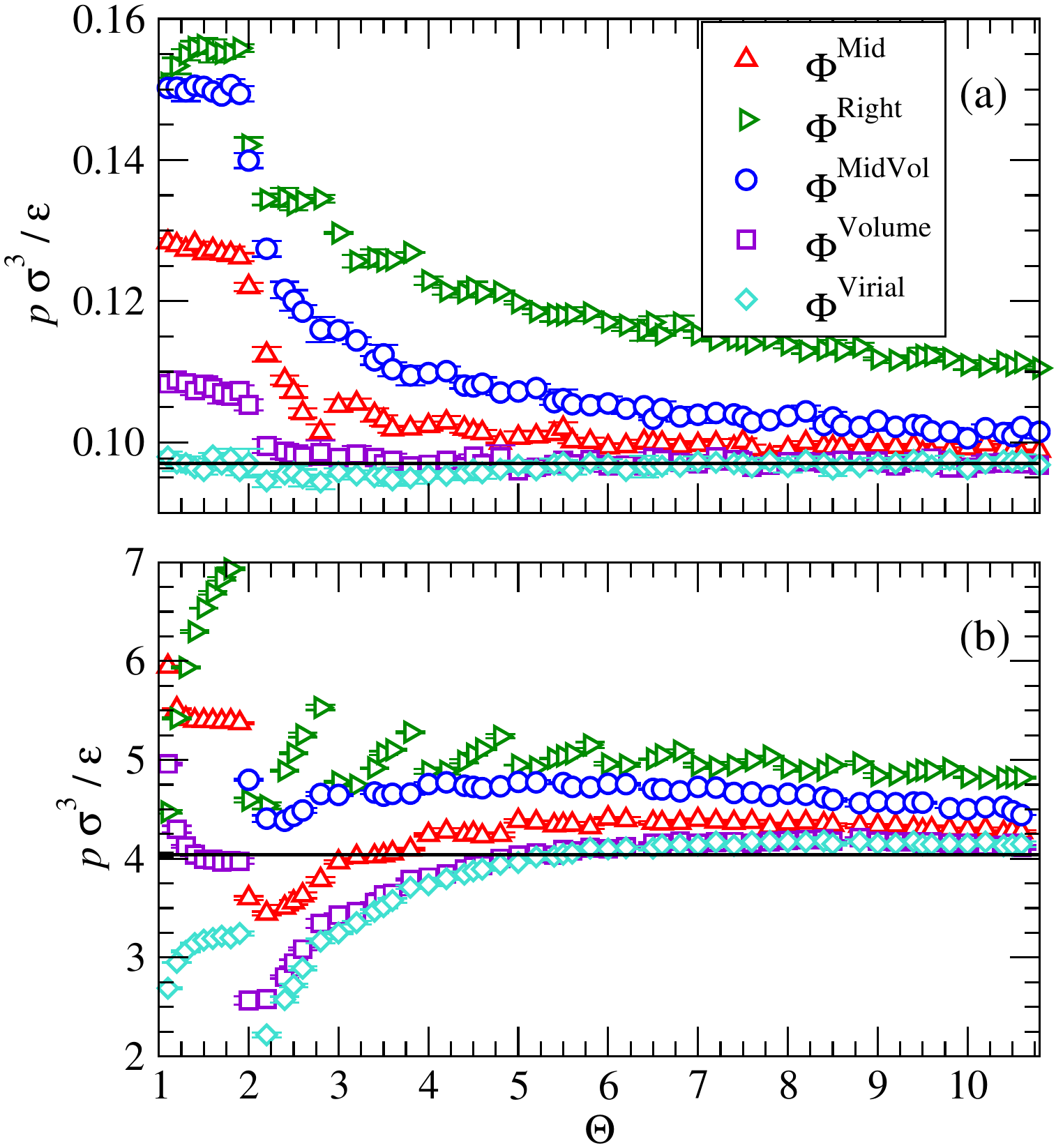}
  \caption{\label{fig:p_phi_comparison} A comparison of the stepped
    potential predictions for the pressure, $p$, of a
    $k_BT/\varepsilon=1.3$ Lennard-Jones fluid as a function of
    attractive step count $\Theta$ at (a) low ($\rho\sigma^3=0.1$) and
    (b) high ($\rho\sigma^3=0.85$) densities. The solid line indicates
    the continuous potential result and each symbol represents a
    different energy stepping algorithm
    (Eqs.~\eqref{eq:LeftEnergyStepping}--\eqref{eq:VirialEnergyStepping}). Discontinuity
    locations are calculated using
    Eq.~\eqref{eq:deltaustepping}. Error bars are the standard
    deviations between the five production runs.}
\end{figure}

\begin{figure}
  \centering
  \includegraphics[width=\columnwidth,clip]{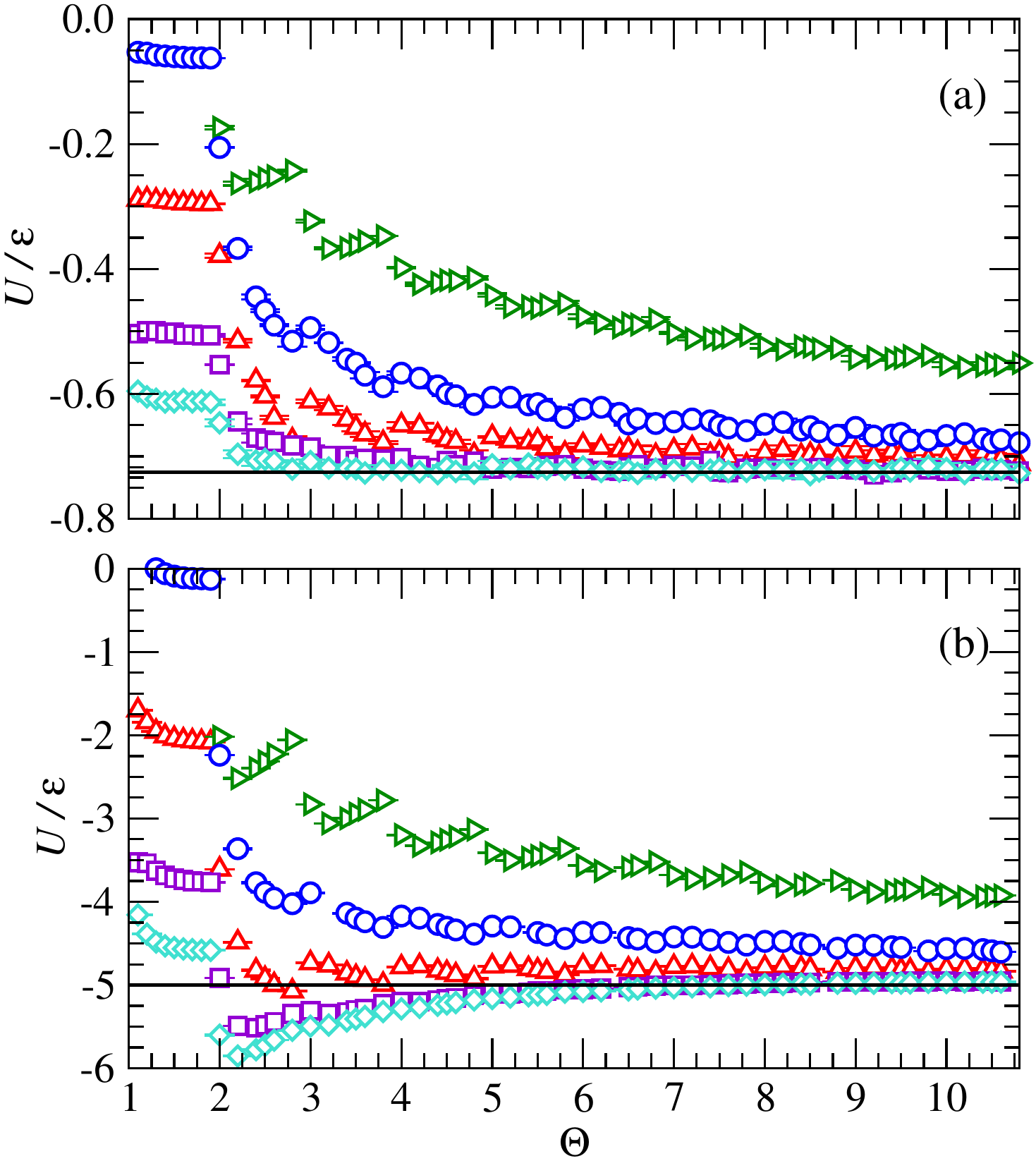}
  \caption{\label{fig:U_phi_comparison} A comparison of the stepped
    potential predictions for the internal energy $U$ for the system
    described in Fig.~\ref{fig:p_phi_comparison}.}
\end{figure}

\begin{figure}
  \centering
  \includegraphics[width=\columnwidth,clip]{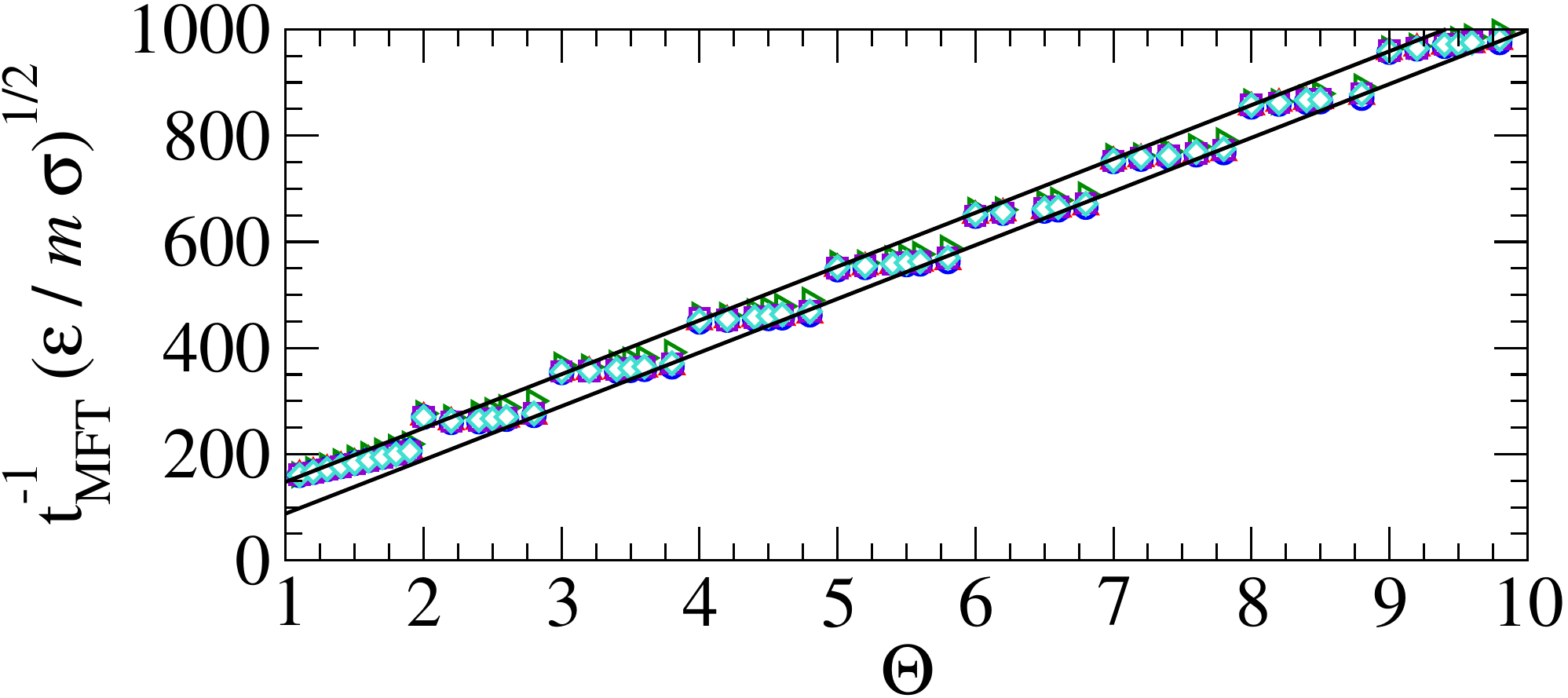}
  \caption{\label{fig:mft_n} The events per particle per unit of
    simulation time, given by $t_{MFT}^{-1} =
    2\,N_{events}/(N\,t_{sim})$ where $N_{events}$ is the number of
    events caused by a particle pair encountering a discontinuity
    during a simulation of duration $t_{sim}$, for various step
    placement algorithms at a temperature of $k_BT/\varepsilon=1.3$
    and density of $\rho\sigma^3=0.85$.  The symbols are described in
    Fig.~\ref{fig:p_phi_comparison} and the straight lines have been
    regressed to the virial data points for $\Theta>2$ with fractional
    parts of $.0$ and $.8$. }
\end{figure}

The algorithms for setting the step energy, given in
Eqs.~\eqref{eq:LeftEnergyStepping}--\eqref{eq:VolumeEnergyStepping},
%and Eq.~\eqref{eq:VirialEnergyStepping}, 
are evaluated by comparing predictions for the pressure (see
Fig.~\ref{fig:p_phi_comparison}) and internal energy (see
Fig.~\ref{fig:U_phi_comparison}) for two super-critical state-points
at low and high density. Equation~\eqref{eq:deltaustepping} is used to
specify the step locations and the order of approximation is again
controlled by specifying the number of discontinuities $\Theta$ in the
attractive section of the potential.  To confirm that $\Theta$ is the
correct basis for comparison of these algorithms, the simulation event
rates as a function of $\Theta$ are presented in
Fig.~\ref{fig:mft_n}. Each step-energy algorithm has an almost
identical cost as a function of $\Theta$ and, for $\Theta>2$, a
remarkable linear correlation is observed between the rate of events.
As the event-driven simulation algorithm processes events at an
approximately constant rate for a given cutoff and density, this
demonstrates that there is a direct correspondence between $\Theta$
and the computational cost, making it a suitable basis for comparison.

%%%
%%%
%%%
%%%

%%%
%%%
%%%
%%%

At low densities, equating the virial contributions provides an
excellent agreement for the pressure for all orders of approximation
(see Fig.~\ref{fig:p_phi_comparison}a), as expected. This is
particularly interesting as for $\Theta\in(1,2)$ the discontinuous
potential is a core-softened square-well potential. For predictions of
the internal energy, the virial approach performs well only once a
step is added between the minimum and the cutoff ($\Theta>2$ in
Fig.~\ref{fig:U_phi_comparison}a). The results of the Left sampling
algorithm are not visible in the figures as they appear to enter the
two-phase region, resulting in a very poor approximation. The Left,
Mid, and Right algorithms display a strong dependence on the step
placement through large changes near integer values of $\Theta$.  The
relatively smooth dependence of the Volume, and Virial algorithms on
the $\Theta$ parameter indicates that these approaches provide a
relatively unbiased sampling of the underlying potential. The MidVol
algorithm performs worse overall than the Mid-point sampling approach
and the Volume averaging algorithm performs almost as well as the
Virial algorithm. The Volume algorithm appears to be a suitable
replacement for the Virial if the temperature is not known and
provided $\Theta>1$. At high densities (see
Figs.~\ref{fig:p_phi_comparison}b and \ref{fig:U_phi_comparison}b),
the Virial and Volume averaging approaches still perform well but
surprisingly the Mid-point sampling outperforms all other techniques
for $1<\Theta<4$. This is likely due to a fortuitous cancellation of
errors rather than an inherent advantage of the method, as its
performance worsens for higher order approximations ($\Theta>4$). It
is clear that an excellent approximation is obtained for $\Theta>5$
with volume or virial stepping.

Overall, setting the step energy through a volume average of the
energy of the underlying continuous potential appears to yield a good
approximation, provided $\Theta>1$. The additional complexity of a
temperature-dependent potential obtained through the Virial approach
does not appear justified at these conditions unless $\Theta<1$, but
the reproduction of the internal energy is unacceptable at such a low
order of approximation. Although it might be expected that the
temperature dependence will become increasingly important at lower
temperatures, further simulations carried out in the liquid branch at
$k_B\,T/\varepsilon=0.85$ and $\rho\sigma^3=0.85$ yielded similar
results to those reported, indicating the temperature correction is
unjustified even when well within the liquid phase.

In summary, the placement of discontinuities using
Eq.~\eqref{eq:deltaustepping} and allocation of their energies using
Eq.~\eqref{eq:VolumeEnergyStepping} 
appears to provide the best 
%allows an optimal and controlled
approximation of the continuous Lennard-Jones potential of the methods
examined for the evaluated state points. As the computational cost
primarily depends on the integer portion of $\Theta$ (see
Fig.~\ref{fig:mft_n}), it is optimal to select values of $\Theta$ with
large fractional parts, such as $\Theta\approx5.8$.

%%%%%%%%%%%%%%%%%%%%%%%%%%%%%%%%%%%%%%%%%%%%%%%%%%%%%%%%%%%%%%%%%%%%%%%%%%%%%%%%
%%%%%%%%%%%%%%%%%%%%%%%%%%%%%%%%%%%%%%%%%%%%%%%%%%%%%%%%%%%%%%%%%%%%%%%%%%%%%%%%
\section{Optimal Algorithm Evaluation
  \label{sec:performance}}

The conversion algorithm which yielded the best performance, given in
Eq.~\eqref{eq:deltaustepping} and Eq.~\eqref{eq:VolumeEnergyStepping},
is now fully evaluated across a wide range of state points. In
particular, the trade-off between accuracy of reproduction and
computational cost is explored. All values of $\Theta$ have a
fractional part of $.8$ due to the step-wise scaling of the
computational cost with $\Theta$ (see Fig.~\ref{fig:mft_n}).

\subsection{Thermodynamic Properties}
\begin{figure}
  \centering
  \includegraphics[width=\columnwidth,clip]{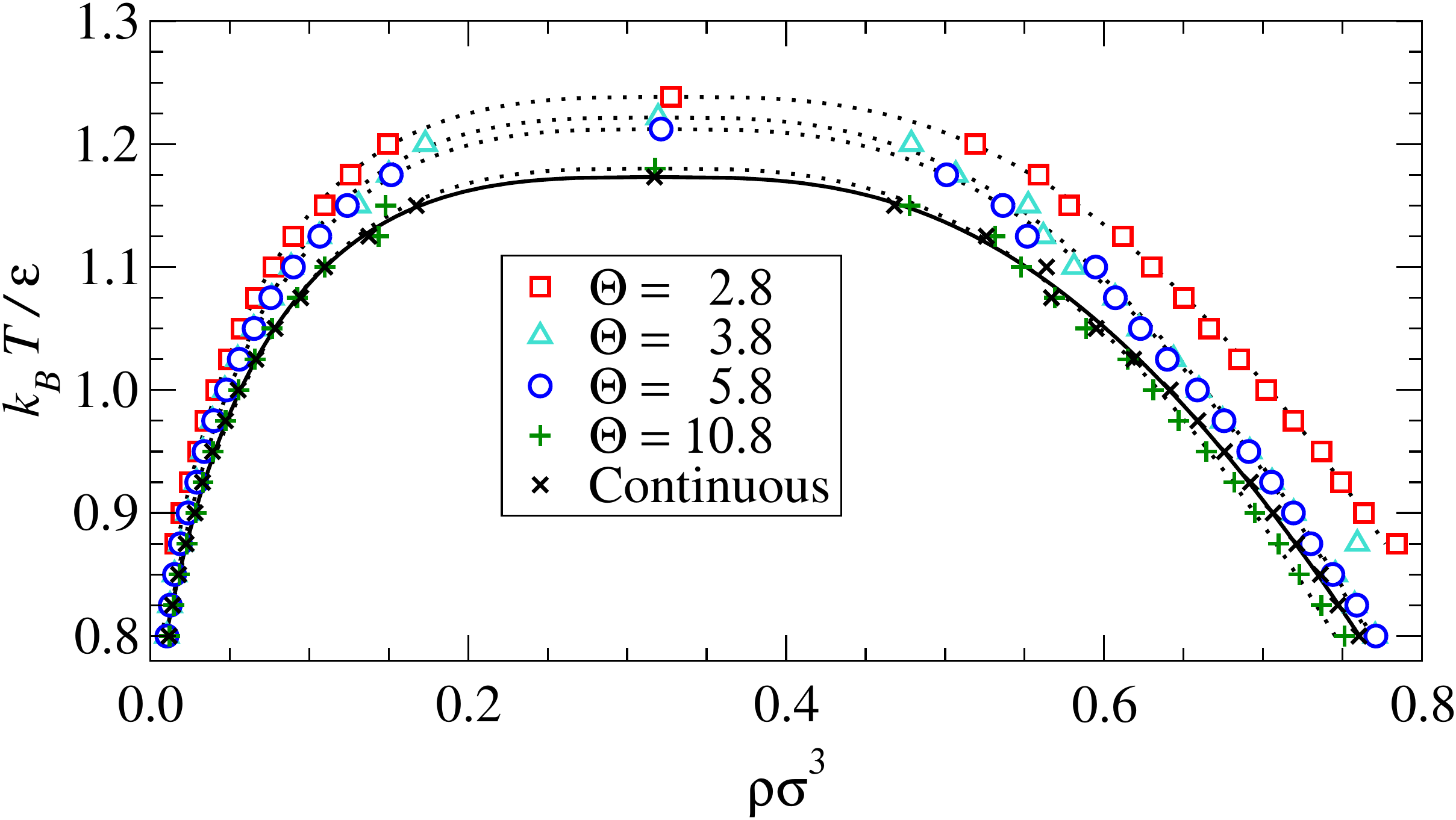}
  \caption{\label{fig:phase_diagram} Phase diagram of the
    $r_\text{cutoff}=3\sigma$ Lennard-Jones fluid. Symbols denote MCMC
    results and the extrapolated critical point. Lines are the best
    fits of Eqs~\eqref{eq:phasefitone} and
    \eqref{eq:phasefittwo}.}
\end{figure}

\begin{figure}
  \centering
  \includegraphics[width=\columnwidth,clip]{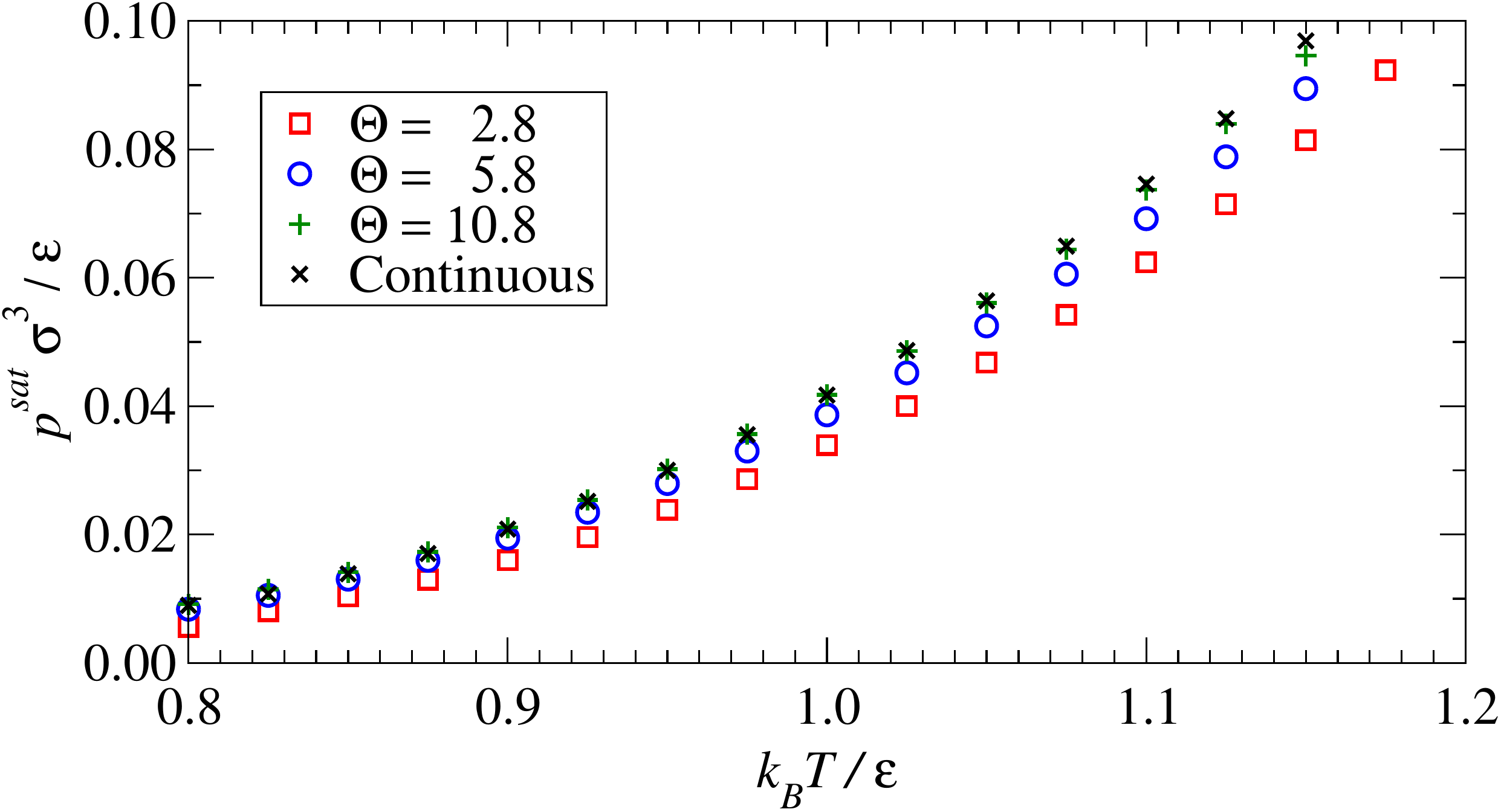}
  \caption{\label{fig:vapourpressure} Vapor pressures $p^{sat}$ of the
    $r_{\text{cutoff}}=3\sigma$ stepped and continuous Lennard-Jones
    fluid.}
\end{figure}

To validate the thermodynamic properties of the system, the phase
diagram (see Fig.~\ref{fig:phase_diagram}) and vapor pressures (see
Fig.~\ref{fig:vapourpressure}) are calculated using Monte Carlo
simulations in the grand canonical ensemble, using multi-canonical
sampling to overcome the free energy barrier between the liquid and
vapor
phases~\cite{Wilding_1995,Orkoulas_Panagiotopoulos_1999,Wilding_2001}.
The simulations were performed in a cubic box of side length
$7\sigma$.  Approximately $100\times10^6$ configurations were sampled
at each temperature, with 50\% attempted displacement moves and 50\%
attempted particle insertion/deletion moves.  Simulations were started
near the critical point, and histogram re-weighting was used to
determine the multi-canonical weights at the lower temperatures.  The
coexistence point at each temperature was determined by adjusting the
chemical potential to equate the areas of the density histogram
corresponding to the liquid and vapor phase.

The critical point is estimated by using a least-squares fit of the
critical scaling of the density difference and the law of rectlinear
diameters
\begin{align}\label{eq:phasefitone}
  \rho_L - \rho_V &= C_1 \left(1-\frac{T}{T_c}\right)^{\beta_c}
  \\ 
  \frac{1}{2}(\rho_L + \rho_V) &= \rho_c + C_2 \left(T-T_c\right) 
\label{eq:phasefittwo}
\end{align}
where $\rho_L$ and $\rho_V$ are the liquid and vapor densities at a
temperature of $T$, and $T_c$ and $\rho_c$ are the critical properties
which, along with $C_1$ to $C_2$, are fitting parameters. A critical
exponent of $\beta_c=0.3265$ is used here.  The optimal conversion
procedure appears to deliver a smooth convergence to the continuous
potential result for the phase envelope in
Fig.~\ref{fig:phase_diagram}, highlighting the value of $\Theta$ as an
order of approximation. The $\Theta=10.8$ system closely reproduces
the thermodynamic behavior of the continuous Lennard-Jones potential,
with only a slight under-estimation of the liquid transition density
for low-temperature liquids.  For values of $\Theta=3.8$ and $5.8$,
the approximation is remarkably close for such a low order
approximation, but for $\Theta=2.8$, the approximation rapidly
deteriorates. Given the relatively low values of $\Theta$, the
potential appears to be performing well when compared to previous
conversions~\cite{Chapela_etal_1989,Chapela_etal_2010,Chapela_etal_2013},
although direct comparisons are difficult due to different choices for
the cutoff radius. Results for the vapor pressures in
Fig.~\ref{fig:vapourpressure} confirm the close reproduction of the
continuous potential phase diagram for the $\Theta=10.8$ system.

%%%
%%%
%%%
%%%

%%%
%%%
%%%
%%%

\begin{figure}
  \centering
  \includegraphics[width=\columnwidth,clip]{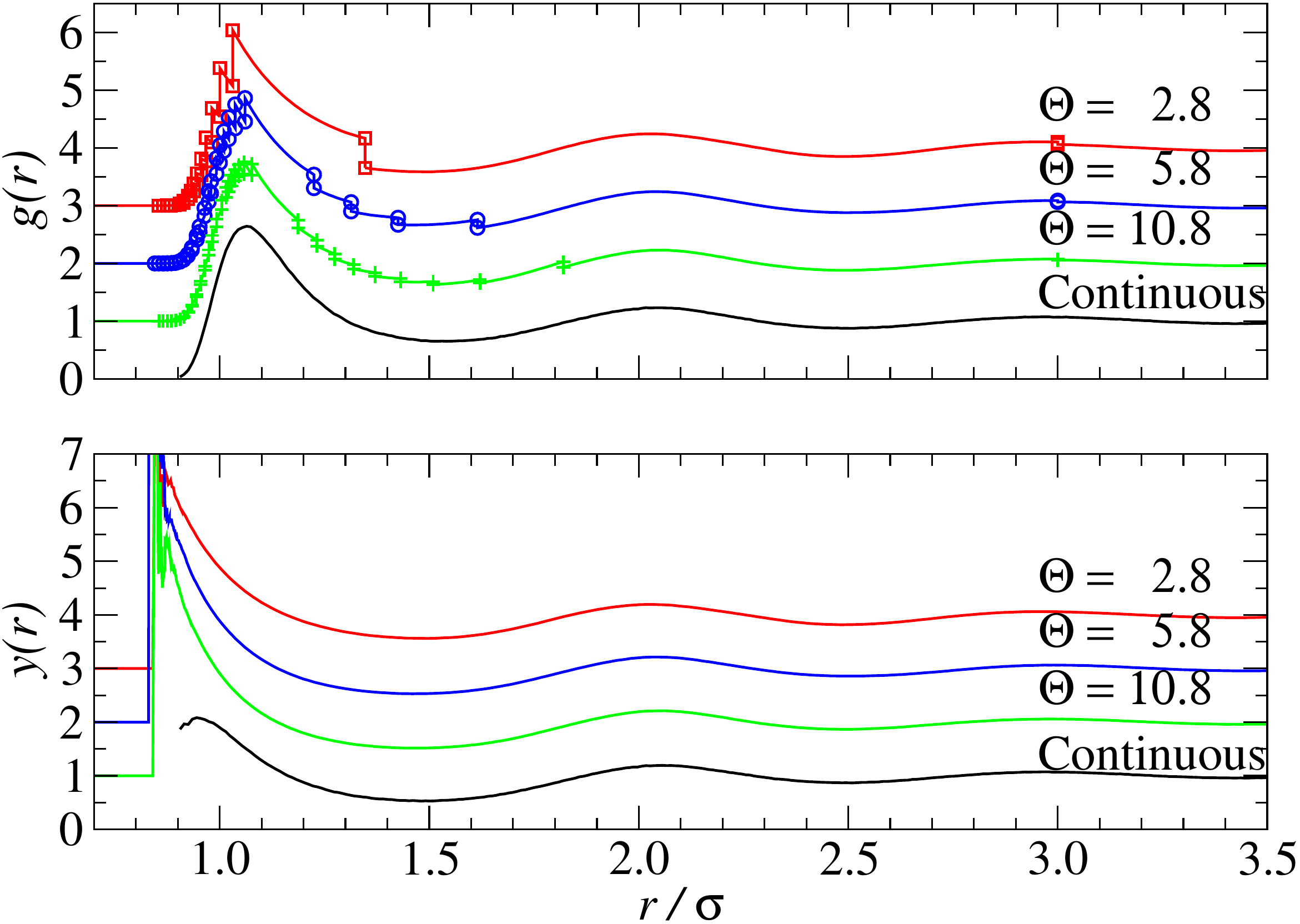}
  \caption{\label{fig:rdf} The (a) radial distribution function,
    $g(r)$, and the (b) cavity correlation function, $y(r)$, for the
    stepped and continuous Lennard-Jones potential at $k_B\,T=1.3$ and
    $N/V=0.85$. Curves have been shifted to ease comparison and
    symbols denote values either side of the discontinuities as
    calculated from the event-rate
    formulas~\cite{Bannerman_Lue_2010}.}
\end{figure}

Although the discrete approximations appear to converge to the
thermodynamic properties of the Lennard-Jones fluid, there are subtle
differences in the microscopic structure. The discontinuities in the
energetic potential lead to discontinuities in $g(r)$, as illustrated
for a high-density state point in Fig.~\ref{fig:rdf}. For low-order
approximations the differences are significant, but for $\Theta=10.8$
the $g(r)$ is closely reproduced. The continuous cavity distribution
function, defined as $y(r)=g(r)e^{\beta\Phi(r)}$, yields a close
agreement between all approximations (see Fig.~\ref{fig:rdf}b). The
sampling of $y(r)$ is poor for the continuous potential near $r\to0.9$
as the $g(r)$ is low in this region; however, the use of the
event-rate formulas allows a higher accuracy of sampling in this
region for the stepped potential which explains the
discrepancy. Despit these small differences in micro-structure, the
overall thermodynamic properties of the system are effectively
captured by the stepped potentials.

%%%
%%%
%%%
%%%

%%%%%%%%%%%%%%%%%%%%%%%%%%%%%%%%%%%%%%%%%%%%%%%%%%%%%%%%%%%%%%%%%%%%%%%%%%%%%%%%%%%%%%%%%
\subsection{Dynamical Properties}

To fully validate the effectiveness of the conversion process, a
number of dynamical properties have been calculated and are compared
against the reference simulation results of Meier et
al.~\cite{Meier_thesis,Meier_etal_2004,Meier_etal_2004b}, and Bugel
and Galliero~\cite{Bugel_Galliero_2008}.  Dynamical properties are
difficult to calculate accurately as they require long simulation
times to allow hydrodynamic behavior to appear. The literature values
used here have been obtained using larger cut-offs (in the range of
$r_\text{cutoff}/\sigma=5$ to $6$, depending on density). This will
only cause small deviations except near the critical point of the
literature fluid ($k_B\,T_c/\varepsilon\approx1.34$) which is above
that of the $r_\text{cutoff}/\sigma=3$ fluid used here
($k_B\,T_c/\varepsilon\approx1.17$). Density sweeps are performed at
two supercritical temperatures, $k_B\,T/\varepsilon=1.35$ near to the
critical point of the reference data and
$k_B\,T/\varepsilon=2.5$. Relative error estimates are obtained by
combining the standard deviation between simulation runs for both the
reference and discontinuous results.
 
\begin{figure}[t]
  \centering
  \includegraphics[width=\columnwidth,clip]{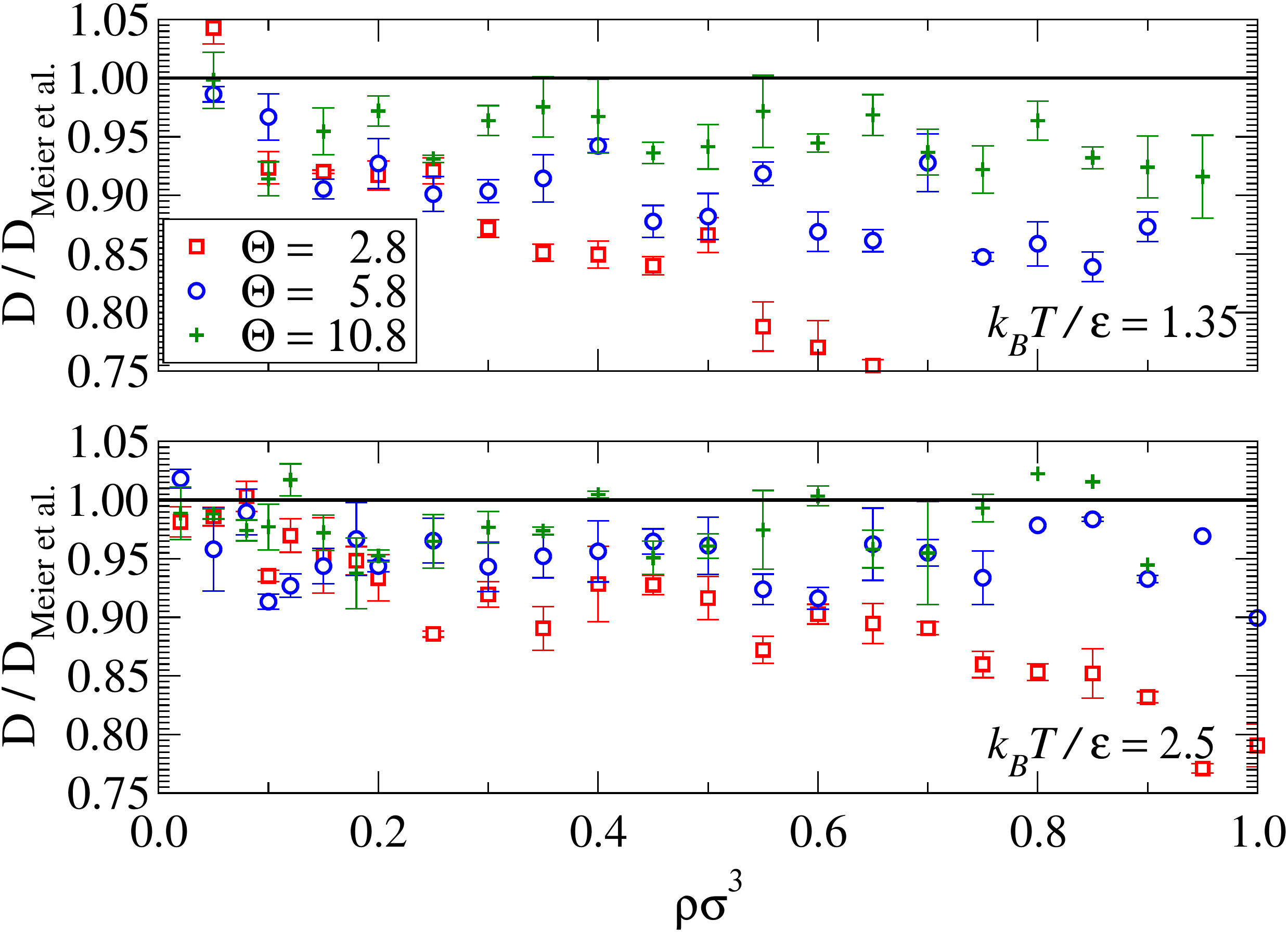}
  \caption{\label{fig:diffusion} Diffusion coefficients relative to
    the reference continuous-potential data for two isotherms over a
    range of densities.  Reference data is taken from Meier et
    al.~\cite{Meier_etal_2004b}.}
\end{figure}

Results for the self-diffusion coefficient are presented in
Fig.~\ref{fig:diffusion}. It is clear that dynamical properties such
as the diffusion coefficient are much more difficult to approximate
when compared with the thermodynamic properties. Low-order
approximations have a reduced diffusion coefficient when compared to
the continuous potential. This effect is due to the difference in
critical temperatures which results in an lower reduced temperature
for the low-order approximations in the theorem of
corresponding-states. The results for $\Theta=10.8$ are acceptable
with a maximum deviation of $\approx8\%$ over both isotherms.

%%%
%%%
%%%
%%%

\begin{figure}
  \centering
  \includegraphics[width=\columnwidth,clip]{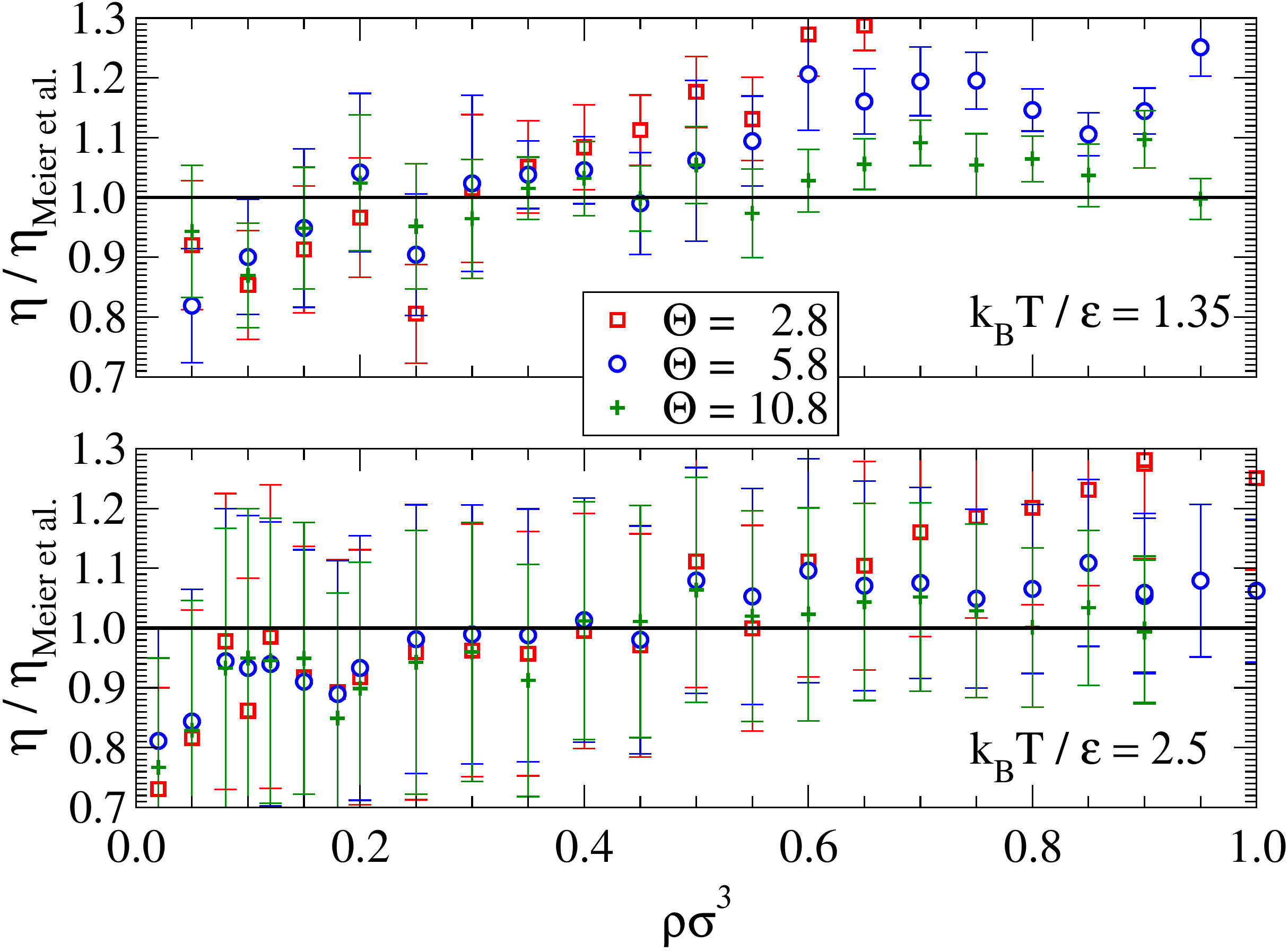}
  \caption{\label{fig:viscosity} Shear viscosity coefficients relative
    to the reference continuous-potential data for two isotherms over
    a range of densities. Reference data is taken from Meier et
    al.~\cite{Meier_etal_2004}.}
\end{figure}

Results for the viscosity are presented in
Fig.~\ref{fig:viscosity}. The $\Theta=10.8$ results are within $10$\%
of the continuous results for both isotherms, except at the lowest two
densities. The disagreement at low densities is likely due to the
increased proportion of ``glancing'' interactions, where the particles
approach each other close enough to pass through the outer attractive
portion of the potential but do not collide directly enough to reach
the inner well or repulsive cores.  The importance of these types of
collisions in this regime shifts the emphasis from a close
reproduction of the well to the outer tail of the potential. It may be
that regular stepping of the potential using radial or volumetric
placement is more suitable at very low gas densities. Overall,
agreement is good, given the uncertainty in the results.

%%%
%%%
%%%
%%%

\begin{figure}
  \centering
  \includegraphics[width=\columnwidth,clip]{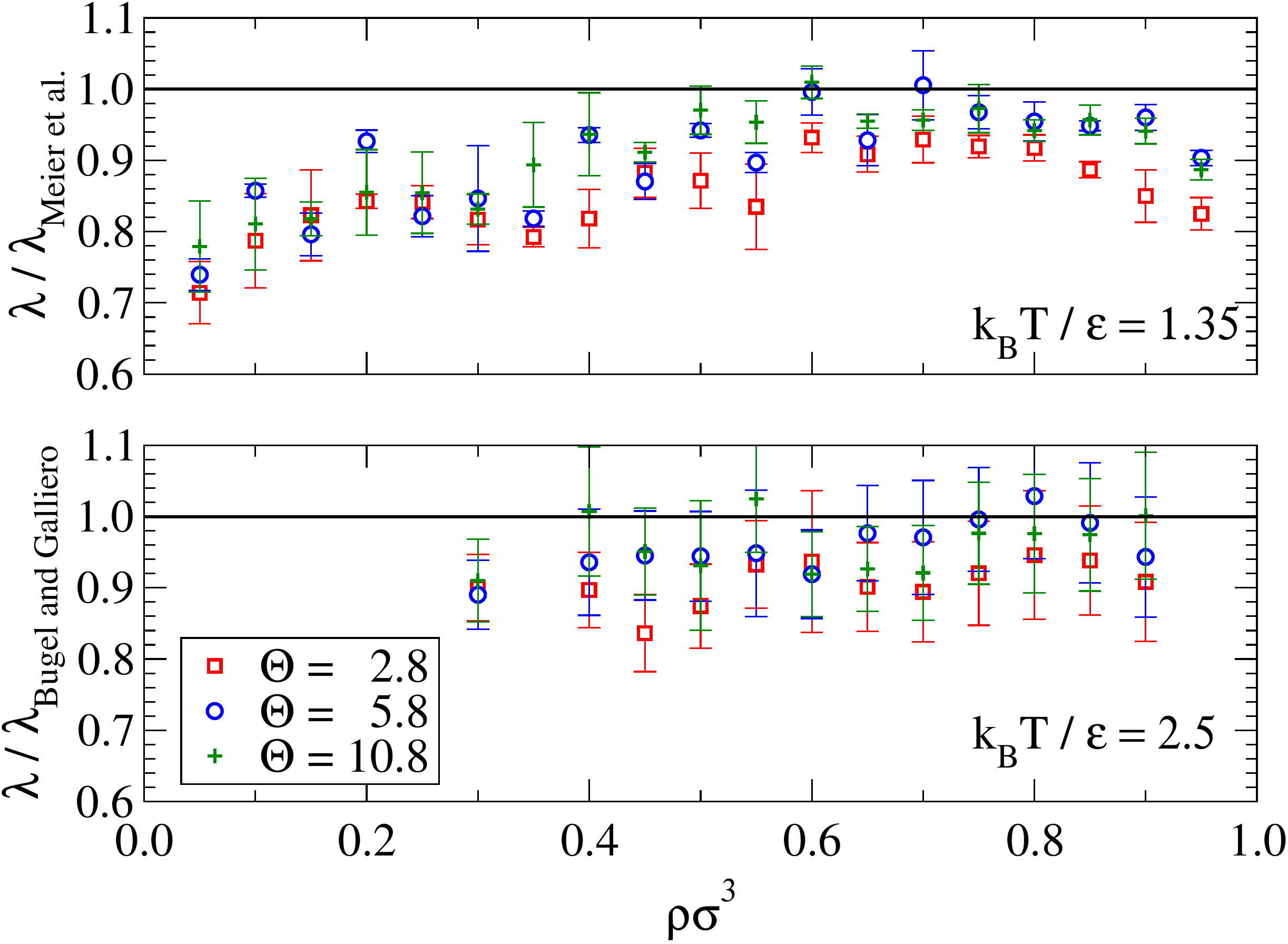}
  \caption{\label{fig:thermalcond} Thermal conductivity relative to
    the reference continuous-potential data for two isotherms over a
    range of densities. Reference data is taken from Meier et
    al.~\cite{Meier_thesis}, and Bugel and
    Galliero~\cite{Bugel_Galliero_2008}.}
\end{figure}

Results for the thermal conductivity are presented in
Fig.~\ref{fig:thermalcond}.  Agreement is good for high temperatures,
but for the $k_B\,T/\varepsilon=1.35$ isotherm the stepped approximations
significantly underestimate the thermal conductivity of the continuous
potential. The disagreement at low-density may again be caused by an
increase in glancing interactions but the discrepancy persists at
moderate densities. Although there is typically very little change in
the thermal conductivity with cutoff range there is an enhancement of
the thermal conductivity in region of the critical
point~\cite{Bugel_Galliero_2008}. As the literature values at this
temperature are near the critical point the stepped potential results
will slightly under predict the thermal conductivity. For liquid
densities ($\rho\,\sigma^3\gtrsim0.6$), the stepped approximation
satisfactorily reproduces the continuous potential behavior.

%%%
%%%
%%%
%%%

%%%%%%%%%%%%%%%%%%%%%%%%%%%%%%%%%%%%%%%%%%%%%%%%%%%%%%%%%%%%%%%%%%%%%%%%%%%%%%%%
\section{\label{sec:compcost} Computational cost}

\begin{figure}
  \centering
  \includegraphics[width=\columnwidth,clip]{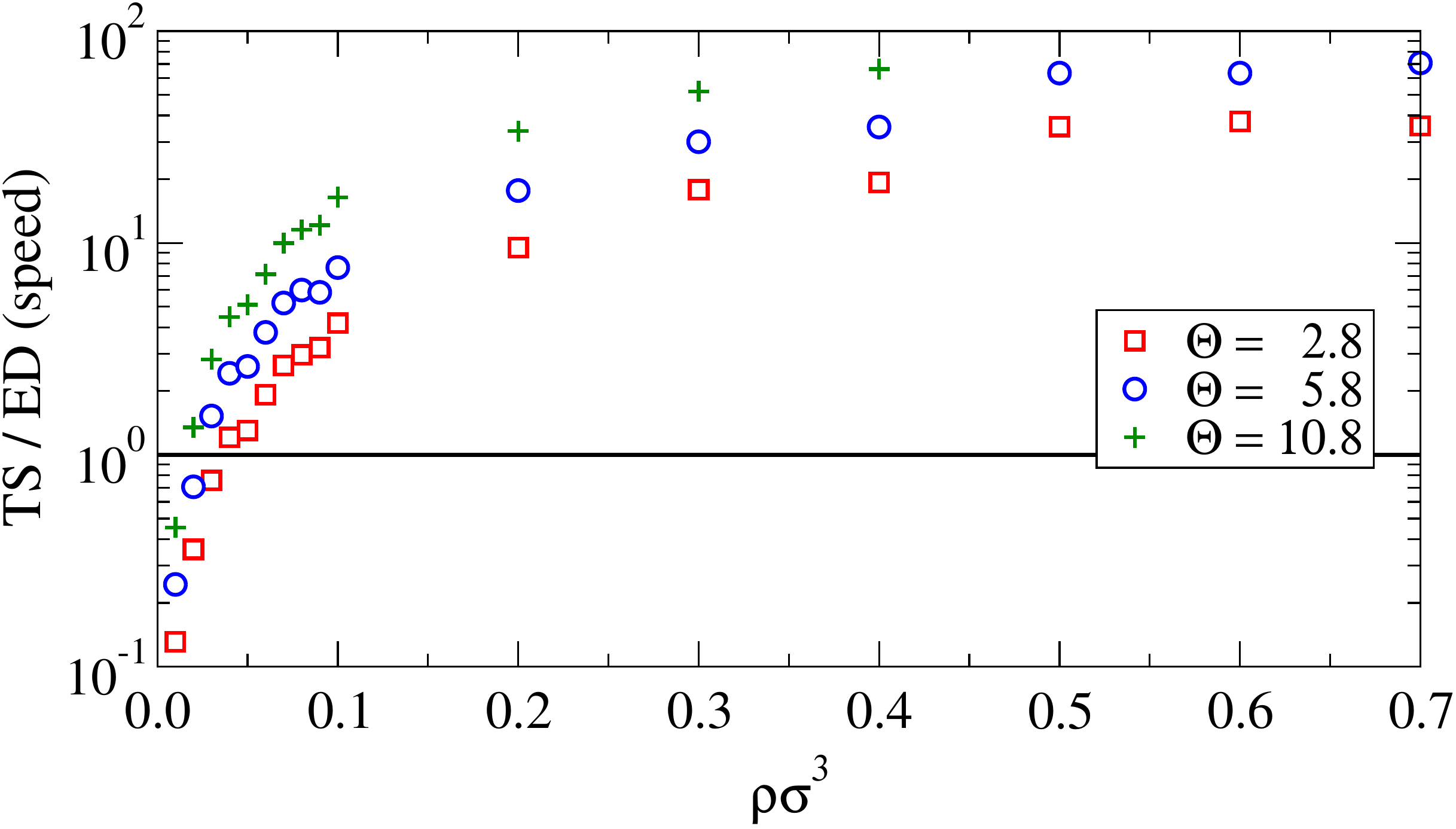}
  \caption{\label{fig:speedup} The relative calculation-speed
    (simulation time processed per unit of CPU time) of time-stepping
    (TS) and event-driven (ED) molecular dynamics for a Lennard-Jones
    isotherm at $k_B\,T/\varepsilon=1.3$. Symbols correspond to
    different stepped approximations.}
\end{figure}

Using the selected conversion procedure, the computational cost of
stepped/event-driven and continuous/time-stepping techniques can be
compared on an equal basis. The relative speed of each method, defined
as the simulation time processed per unit of CPU time, for a
$k_B\,T/\varepsilon=1.3$ isotherm is presented in
Fig.~\ref{fig:speedup}. It is immediately apparent that, for the
Lennard-Jones potential and the software packages considered here, the
use of event-driven methods is only advantageous at gas densities
($\rho\,\sigma^3\lesssim0.02$) or below, where it significantly
outperforms time stepping methods. Hybrid time-stepping/event-driven
methods also display this dramatic increase in performance at low
densities~\cite{Valentini_Schwartzentruber_2007} as the system
dynamics becomes dominated by two-particle collisions. The use of
stepped potentials will eliminate some of the overhead of hybrid
techniques which indicates promising future applications of the
potentials developed here in rarefied gas flow simulation.  The
comparison carried out here is only for serial execution performance;
however, parallel algorithms for event-driven simulation exhibit good
scaling~\cite{Miller_Luding_2004} and will be evaluated in future.

%% The hybrid paper is performing simulations from 0.0004 to 0.0000004
%% density!

%% The crossover point in performance is around 33 kg/m^3 for Argon.

%%%
%%%
%%%
%%%

%%%%%%%%%%%%%%%%%%%%%%%%%%%%%%%%%%%%%%%%%%%%%%%%%%%%%%%%%%%%%%%%%%%%%%%%%%%%%%%%
%%%%%%%%%%%%%%%%%%%%%%%%%%%%%%%%%%%%%%%%%%%%%%%%%%%%%%%%%%%%%%%%%%%%%%%%%%%%%%%%
\section{Conclusions \label{sec:conclusions}}

In this work, we have examined several methods for mapping a
continuous interaction potential to a discrete, stepped potential.
These methods were compared on their ability of the resulting discrete
potential to reproduce the thermodynamic and transport properties of
the continuous potential system over a broad range of conditions.  Of
the various methods which were examined to locate the discontinuities
of the potential, the best was found to be at fixed intervals of
energy.  Setting the step energy through a volume average of the
energy of the underlying continuous potential appears to give the a
good overall approximation provided $\Theta>1$ and an excellent
approximation for $\Theta=10.8$.

The computational cost for simulating the continuous potential and
several levels of stepped potential approximation were compared on the
basis of their processing rates for a unit of simulation time.
%
%The comparison between the event-driven and continuous molecular
%dynamics methods is made on the basis of the CPU time required to
%simulate a set length of the system time, 
%
This is dependent on the particular algorithms used to perform each of
the simulations, and, consequently, the implementation of different
algorithms may alter the results of this comparison.
Stepped Lennard-Jones potentials are increasingly efficient when
compared to continuous forms at gas densities or lower
($\rho\,\sigma^3\lesssim0.02$ which is around $33$\,kg\,m${}^{-3}$ for
Argon). This indicates the stepped potentials introduced here may be
used to significantly accelerate simulations of shock
waves~\cite{Valentini_Schwartzentruber_2007} or other complex rarefied
gas systems where DSMC techniques are currently applied.

In our comparisons, we have only studied the use of stepped potentials
to reproduce the properties of a system where the particles interact
with a \emph{continuous} potential.  In that respect, continuous
molecular dynamics algorithms are at a distinct advantage as, in order
for discrete potentials to properly reproduce the properties of a
system interacting with a continuous potential, a fairly large number
of steps is required. Models with shorter interaction durations, such
as the Weeks-Chandler-Anderson potential, should prove far more
efficient targets for conversion, especially given the speed of
hard-sphere simulations. The Hertz potential, used in simulations of
solids particles~\cite{Poschel_Schwager_2005}, is particularly
interesting as the stepped equivalent may be arbitrarily steep. This,
combined with the analytic dynamics of the event-driven technique,
will allow the stable and efficient use of realistic materials
parameters in granular simulations. The only obstacle to this
application is the conversion of the dissipative inter-particle forces
which will be explored in a future publication.

Finally, in this paper, we have not considered the general question as
to whether using a discrete potential is preferable to using a
continuous potential for molecular systems.  Although the discrete
Lennard-Jones model developed here only appears to be efficient in
certain limits, using a carefully chosen discrete coarse-grained
potential, with only a few steps and a short interaction range, can
potentially lead to a system that is much more rapidly simulated by
event-driven MD than a continuous potential can be simulated by
continuous MD.  For example, the use of discrete interaction
potentials, along with a combination of event-driven MD and
thermodynamic perturbation theory, have greatly improved the speed of
developing and tuning transferable potential models to accurately
reproduce the thermodynamic properties and phase behavior of fluids
\cite{Ucyigitler_etal_2012}.  Also, using discrete potential models
has allowed the rapid simulation and large, complex biological systems
\cite{Dokholyan_2006}, such as the dynamics of protein folding
\cite{Ding_etal_2005,Ding_etal_2005b} and fibril formation
\cite{Cheon_etal_2011}, over long time scales.  These types of
simulations are still difficult to access by current continuous
potential techniques and highlight the potential of carefully
developed discontinuous models.

%%%%%%%%%%%%%%%%%%%%%%%%%%%%%%%%%%%%%%%%%%%%%%%%%%%%%%%%%%%%%%%%%%%%%%%%%%%%%%%%
%%%%%%%%%%%%%%%%%%%%%%%%%%%%%%%%%%%%%%%%%%%%%%%%%%%%%%%%%%%%%%%%%%%%%%%%%%%%%%%%
\section{Acknowledgments}

The authors would like to acknowledge the support of the Maxwell
compute cluster funded by the University of Aberdeen.

%%%%%%%%%%%%%%%%%%%%%%%%%%%%%%%%%%%%%%%%%%%%%%%%%%%%%%%%%%%%%%%%%%%%%%%%%%%%%%%%
%%%%%%%%%%%%%%%%%%%%%%%%%%%%%%%%%%%%%%%%%%%%%%%%%%%%%%%%%%%%%%%%%%%%%%%%%%%%%%%%

%merlin.mbs aipnum4-1.bst 2010-07-25 4.21a (PWD, AO, DPC) hacked
%Control: key (0)
%Control: author (8) initials jnrlst
%Control: editor formatted (1) identically to author
%Control: production of article title (-1) disabled
%Control: page (0) single
%Control: year (1) truncated
%Control: production of eprint (0) enabled
%
%%%%%%%%%%%%%%%%%%%%%%%%%%%%%%%%%%%%%%%%%%%%%%%%%%%%%%%%%%%%%%%%%%%%%%%%%%%%%%%%
%%%%%%%%%%%%%%%%%%%%%%%%%%%%%%%%%%%%%%%%%%%%%%%%%%%%%%%%%%%%%%%%%%%%%%%%%%%%%%%%
%%%%%%%%%%%%%%%%%%%%%%%%%%%%%%%%%%%%%%%%%%%%%%%%%%%%%%%%%%%%%%%%%%%%%%%%%%%%%%%%
\end{document}